\documentclass[letterpaper,superscriptaddress, aps, prc, reprint, amsmath,amssymb]{revtex4-1}
\usepackage{graphicx} 

\begin{document}

\title{Robust extraction of the proton charge radius from electron-proton scattering data}

\author{Xuefei Yan}
\email{xy33@phy.duke.edu}
\affiliation{Duke University, Durham, North Carolina 27708, USA}
\affiliation{Triangle Universities Nuclear Laboratory, Durham, North Carolina 27708, USA}

\author{Douglas W. Higinbotham}
\affiliation{Thomas Jefferson National Accelerator Facility, 12000 Jefferson Avenue, Newport News, Virginia 23606, USA}

\author{Dipangkar Dutta}
\affiliation{Mississippi State University, Mississippi State 39762, USA}

\author{Haiyan Gao}
\affiliation{Duke University, Durham, North Carolina 27708, USA}
\affiliation{Triangle Universities Nuclear Laboratory, Durham, North Carolina 27708, USA}
\affiliation{Duke Kunshan University, Jiangsu 215316, China}

\author{Ashot Gasparian}
\affiliation{North Carolina A\&T State University, Greensboro, North Carolina 27411, USA}

\author{Mahbub A. Khandaker}
\affiliation{Idaho State University, Idaho 83209, USA}

\author{Nilanga Liyanage}
\affiliation{University of Virginia, Charlottesville, VA 22904, USA}

\author{Eugene Pasyuk}
\affiliation{Thomas Jefferson National Accelerator Facility, 12000 Jefferson Avenue, Newport News, Virginia 23606, USA}

\author{Chao Peng}
\affiliation{Duke University, Durham, North Carolina 27708, USA}
\affiliation{Triangle Universities Nuclear Laboratory, Durham, North Carolina 27708, USA}

\author{Weizhi Xiong}
\affiliation{Duke University, Durham, North Carolina 27708, USA}
\affiliation{Triangle Universities Nuclear Laboratory, Durham, North Carolina 27708, USA}

\date{\today}

\begin{abstract}
\begin{description}
\item[Background] Extracting the proton charge radius from electron scattering data,
requires determining the slope of the charge form factor at $Q^2$ of zero.    
As experimental data cannot reach that limit, numerous methods for making the extraction 
have been proposed.
\item[Purpose] In this study, we seek to find functional forms that will allow a robust extraction 
of the proton radius from a wide variety of functional forms.  The primary motivation of this study is to 
have confidence in the extraction of upcoming low $Q^2$ experimental data.
\item[Method] We create a general framework for studying various form-factor functions
along with various fitting functions.   
The input form factors are used to generate pseudo-data with fluctuations
mimicking the binning and random uncertainty of a set of real data.  
All combinations of input functions and fit functions
can then be tested repeatedly against regenerated pseudo-data.   Since the input radius is known,
this allows us to find fitting functions that are robust for proton radius extractions in an objective fashion.
\item[Results] For the range and uncertainty of the PRad data, we find that a two-parameter rational
function, a two-parameter continued fraction and the second order polynomial expansion of $z$ can extract 
the input radius regardless of the input charge form factor function that is used.    
\item[Conclusions]
We have created a framework to determine which functional forms allow for a robust 
extraction of the radius from pseudo-data generated from a wide variety of trial functions.
By taking into account both bias and variance, the optimal functions for extracting the
proton radius can be determined.
\end{description}
\end{abstract}

\maketitle

%
%
\section{\label{introd}Introduction}

Much effort has been devoted to the determination of the charge radius of the proton ($R$), 
but results from different experiments and/or analyses exhibit sizable discrepancies.
For example, in high-precision muonic hydrogen Lamb shift experiments, $R$ was measured 
to be $0.8409 \pm 0.0004$ fm~\cite{Pohl_2010, Antognini_1900n}, while the current value from 
CODATA, determined from atomic Lamb shift and electron-proton ($ep$) scattering experiments,
is $R = 0.8751 \pm 0.0061$ fm~\cite{Mohr:2015ccw}.
This difference is known as the proton radius puzzle~\cite{Pohl_2013yb, Carlson_2015jba,Hill:2017wzi}.
The newer atomic Lamb shift and electron scattering results that have become
available~\cite{Mihovilovic:2016rkr,Beyer79,Fleurbaey:2018} thus far are contradictory and 
the proton radius puzzle remains.

To extract the proton radius from ep-scattering data the electric form factor, $G_E$, 
is first plotted as a function of the four-momentum transferred squared, $Q^2$.
This data must then be fit to find the slope at $Q^2 = 0$.   The radius depends on the
slope according to Eq.~\ref{eq:slope}. 
Since experimental electron scattering cannot reach the $Q^2 = 0$ limit, 
many different methods have been proposed to extract the radius from the
data.

\begin{equation}
   R \equiv \left( -6  \left. \frac{\mathrm{d} G_E(Q^2)}{\mathrm{d}Q^2}
    \right|_{Q^{2}=0} \right)^{1/2} \>
\label{eq:slope}
\end{equation}

Recent global analyses of $ep$-scattering data 
found $R \approx 0.84$ fm, in agreement with the muonic Lamb shift 
results~\cite{Belushkin:2006qa, Lorenz:2012tm, Lorenz:2014vha, Lorenz:2014yda, Griffioen_2015hta, 
Higinbotham_2015rja, Horbatsch_2015qda, Horbatsch:2016ilr}.
Though these analyses used existing experimental data, they systematically extract smaller radii 
than the results of other 
groups~\cite{Arrington_2015ria, Lee_2015jqa, Borisyuk_2009mg, Graczyk_2014lba, Bernauer_2010wm, Bernauer_2013tpr}.
It has been pointed out that the difference between the results is mainly due to differences in how the
high-order moments $\langle r^{2n} \rangle$ ($n > 1$) are handled~\cite{Sick_2017aor,Sick:2018fzn}.   
A summary table of the higher order moments from a number of these fits can be found in the 
recent work of Alarc{\'o}n and Weiss~\cite{Alarcon:2017lhg}.

The form-factor $G_E$ is often fit with a multi-parameter polynomial expansion of $Q^2$ up to an order $Q^{2N}$,
since each moment $\langle r^{2n} \rangle$ ($1 \leq n \leq N$) corresponds to an 
independent parameter.   Though this description seems to be model independent, as Kraus~\textit{et al.} have shown, 
it does not ensure a correct $R$ extraction when it is used for extrapolating beyond the data 
to $Q^2 = 0$~\cite{Kraus_2014qua}.

In addition to multi-parameter polynomials, functional forms of $G_E$ 
based on models of the proton charge distribution are also used to determine R. 
The problem with this approach is that it can be difficult to quantify how much the
extraction of $R$ is affected by the assumptions in the model.
In addition, constructing a model description of the full charge distribution of the proton is a far more complex problem
than simply trying to mathematically extract $R$ value from experimental data.

Herein we present a systematic method to find mathematical function(s) that 
can robustly extract $R$ over a broad set of $G_E$ input functions.   
In this study, we use  the expected binning and uncertainty of the PRad experiment~\cite{Gasparian:2014rna, Peng:2016szv}
as an example, but the method can be applied to any expected binning and uncertainty.  

%
%
\section{\label{method}Method}

If the exact functional form of the proton's charge form factor, $G_E$, were available, one could fit 
experimental data to this same functional form and extract the charge radius.
This ideal case is easily simulated by creating randomized pseudo-data and examining the fitting results $R(\textrm{fit})$ 
using the same functional form as was used to generate the pseudo-data. 
This process can be repeated multiple times in order to obtain a distribution of $R(\textrm{fit})$. 

However, as the true functional form is unknown, one has to search for functional
forms that can extract R by extrapolating to $Q^2 = 0$ from experimental data.
For simplicity, we call this feature robustness.
In fact, due to the variability of experimental data, the best fitting function
may not even be the true functional form~\cite{Shmueli:2010}.

To find appropriate functions for a given binning and uncertainty, we generate pseudo-data using 
a wide variety of functional forms.   Next, we systematically fit each set of pseudo-data with 
various functional forms.
By studying the distributions of the results, we find functional forms that robustly
extract the input radius.  To be considered robust, the set of extracted R values must be,
within errors, the same as the R value used to generate the pseudo-data regardless of 
which $G_E$ parameterization was used in the generating function.

A program library has been built with three parts to generate pseudo data, add fluctuations,
and fit the pseudo data~\cite{Radius_fitting_lib}. 
This program library is coded in C++ using the Minuit and CERN ROOT package~\cite{Brun:1997pa,James:1975dr}. 
The three components of this library are described in detail in the following subsections.

\subsection{\label{generator}Generator}

The generator library has been built to generate $G_E$ values at given $Q^2$ using either
simple standard functions, parameterizations of experimental data or full theoretical calculations.
Other functions could easily be added to this library.   The currently installed functions include:

\paragraph{Dipole}
The dipole functional form of $G_E$ \cite{Borkowski1975} is expressed as
\begin{eqnarray}
G_E(Q^2) &=& \left (1+\frac{Q^2}{p_1} \right )^{-2} , \label{dipole-ge-eq}
\end{eqnarray}
where $p_1 = 12 / R^2$.  This functional form corresponds 
to an exponential charge distribution of the proton, and the relation between moments is
\begin{eqnarray}
\langle r^{2n} \rangle &=& \frac{(n+1) (2n+1)}{6} \langle r^{2} \rangle \langle r^{2n-2} \rangle , \label{dipole-mom-eq}
\end{eqnarray}
where $n > 1$.

\paragraph{Monopole}
The monopole functional form of $G_E$ \cite{Borkowski1975} is expressed as
\begin{eqnarray}
G_E(Q^2) &=& \left (1+\frac{Q^2}{p_1} \right )^{-1} , \label{monopole-ge-eq}
\end{eqnarray}
where $p_1 = 6 / R^2$. This functional form corresponds to a Yukawa charge distribution of the proton, 
and the relation between moments is
\begin{eqnarray}
\langle r^{2n} \rangle &=& \frac{n(2n+1)}{3} \langle r^{2} \rangle \langle r^{2n-2} \rangle , \label{monopole-mom-eq}
\end{eqnarray}
where $n > 1$.

\paragraph{Gaussian}
The Gaussian functional form of $G_E$ \cite{Borkowski1975} is expressed as
\begin{eqnarray}
G_E(Q^2) &=& \exp(-Q^2/p_1) , \label{gaussian-ge-eq}
\end{eqnarray}
where $p_1 = 6 / R^2$. This functional form corresponds to a Gaussian charge distribution of the proton, 
and the relation between moments is
\begin{eqnarray}
\langle r^{2n} \rangle &=& \frac{2n+1}{3} \langle r^{2} \rangle \langle r^{2n-2} \rangle , \label{guassian-mom-eq}
\end{eqnarray}
where $n > 1$.

\paragraph{Kelly-2004}
The parameterization from Ref. \cite{Kelly_2004hm} is expressed as
\begin{eqnarray}
G_E(Q^2) &=& \frac{1+a_1 \tau}{1+b_1 \tau +b_2 \tau^2 +b_3 \tau^3} , \label{kelly-ge-eq}
\end{eqnarray}
where $\tau = Q^2/4 m_p^2$, and $m_p$ is the proton mass. 
The parameters $a_1$, $b_1$, $b_2$ and $b_3$ can be found in Table I of Ref. \cite{Kelly_2004hm}.
The radius in this parameterization is $R = 0.8630$ fm.

\paragraph{Arrington-2004}
The parameterization from Ref. \cite{Arrington_2003qk} is expressed as
\begin{eqnarray}
G_E(Q^2) &=& \left(1+\sum\limits_{i=1}^N p_{2i} Q^{2i} \right)^{-1} , \label{arrington-ge-eq}
\end{eqnarray}
where parameters $p_{2i}$ up to $i = 6$ can be found in Table I of Ref. \cite{Arrington_2003qk}. 
The radius in this parameterization is $R = 0.8682$ fm.

\paragraph{Arrington-2007}
The parameterization from Ref. \cite{Arrington:2006hm} is a fifth-order continued-fraction (CF) expansion expressed as: 
\begin{eqnarray}
G_E(Q^2) &=& \frac{1}{1+\frac{p_1 Q^2}{1+\frac{p_2 Q^2}{1+ \cdots}}} , \label{cf-as}
\end{eqnarray}
where the parameters $p_i$ (index $i$ from 1 to 5) can be found in Table I in Ref.~\cite{Arrington:2006hm}. 
The radius in this parameterization is $R = 0.8965$ fm.

\paragraph{Venkat-2011}
The parameterization from Ref. \cite{Venkat_2010by} is expressed as
\begin{eqnarray}
G_E(Q^2) &=& \frac{1+a_1 \tau+a_2 \tau^2 +a_3 \tau^3}{1+b_1 \tau +b_2 \tau^2 +b_3 \tau^3+b_4 \tau^4 +b_5 \tau^5} , 
\label{venkat-ge-eq}
\end{eqnarray}
where parameters $a_i$ and $b_i$ can be 
found in Table II of Ref.  \cite{Venkat_2010by}. The radius in this parameterization is $R = 0.8779$ fm.

\paragraph{Bernauer-2014}
This parameterization is a refit of the full set of 1422 data points from Ref.~\cite{Bernauer_2013tpr}
and is expressed as a 10th-order polynomial expansion of $Q^2$:
\begin{eqnarray}
G_E(Q^2) &=& 1 + \sum\limits_{i=1}^{10} p_{i} Q^{2i} , \label{bernauer-2010}
\end{eqnarray}
where the refitted parameters $p_i$ are close to those found in appendix J.1 of Ref.~\cite{JB_thesis}.
The radius in this parameterization is $R = 0.8868$ fm.

\paragraph{Alarc{\'o}n-2017}
As a fully realistic charge form factor, we used the model of
Alarc{\'o}n and Weiss~\cite{Alarcon:2017ivh,Alarcon:2017lhg,Alarcon:2018irp} referred to
herein as Alarc{\'o}n-2017.
This model uses the recently developed method combining chiral effective field theory 
and dispersion analysis.   Solely for the purpose of testing extraction techniques,
the radius in the model was fixed to a series of values: 0.84 fm from muonic
hydrogen, 0.875 fm from CODATA, and 0.85 fm as the central value from the range of radii
allowed by the model.   Unlike the other models where a simple function could be programmed,
here we have used a finely spaced table of charge values and then fit it with a cubic spline.  The
spline function can then be called in a similar manner to the other functions.  

\paragraph{Ye-2018}
The parameterization of Ye~{\it{et al.}}~\cite{Ye:2017gyb} is a fit to world
data with the radius fixed to $R = 0.879$ fm. 
The parameterization and the values of the parameters can be found in the supplemental
materials of Ref.~\cite{Ye:2017gyb}.
The author Z.~Ye also provided a separate parameterization with a different fixed radius, $R = 0.85$ fm. 
This second parameterization will be referred to as Ye-2018~(re-fix) in this study.

\subsection{\label{fluc_adder}Fluctuation-adder}
In order to mimic the variability of real data, library allows adding bin-by-bin and/or
overall fluctuations to the $G_E$ vs. $Q^2$ tables. 
It includes fluctuations according to a user-defined random Gaussian distribution, $\mathcal{N}(\mu, \, \sigma_g^2)$.
In the bin-by-bin case, the uncertainty $\delta G_E$ of each bin is defined by the user. 
The library sets $\mu = 0$ and $\sigma_g = \delta G_E$, 
and generates fluctuations according to $\mathcal{N}(\mu, \, \sigma_g^2)$ in each bin.
In the overall case, the user can manually set the values of $\mu$ and $\sigma_g$, and the 
library generates an overall scaling factor according to $\mathcal{N}(\mu, \, \sigma_g^2)$ 
for all the bins in a table.   Other types of fluctuations, such as uniform and Breit-Wigner,
are also included in the library for test purposes.

\subsection{\label{fitter}Fitter}
To study which function robustly extract R from the generated pseudo-data,
a fitting routine has been developed.
This library uses the Minuit package of CERN ROOT to fit the $G_E$ vs. $Q^2$ tables 
with the functional forms listed below:

\paragraph{Dipole}
The dipole fitter is expressed as
\begin{eqnarray}
f_{\textrm{dipole}}(Q^2) &=& p_0 G_E(Q^2) = p_0 \left (1+\frac{Q^2}{p_1} \right )^{-2} , \label{dipole-ge-eq-fit}
\end{eqnarray}
where $p_0$ is a floating normalization parameter, and $p_1$ is a fitting parameter related to the radius $R = \sqrt{12/p_1}$. 

\paragraph{Monopole}
The monopole fitter is given by
\begin{eqnarray}
f_{\textrm{monopole}}(Q^2) &=& p_0 G_E(Q^2) = p_0 \left (1+\frac{Q^2}{p_1} \right )^{-1} , \label{monopole-ge-eq-fit}
\end{eqnarray}
and $R = \sqrt{6/p_1}$.

\paragraph{Gaussian}
The Gaussian fitter has the form
\begin{eqnarray}
f_{\textrm{Gaussian}}(Q^2) &=& p_0 G_E(Q^2) = p_0 \exp(-Q^2/p_1) , \label{gaussian-ge-eq-fit}
\end{eqnarray}
and $R = \sqrt{6/p_1}$.

\paragraph{Multi-parameter polynomial-expansion of $Q^2$}
The fitter of the multi-parameter polynomial-expansion of $Q^2$ is written as
\begin{eqnarray}
f_{\textrm{polyQ}}(Q^2) &=& p_0 G_E(Q^2) = p_0 \left(1 + \sum\limits_{i=1}^N p_{i} Q^{2i} \right) , \label{poly-Q2-expan-fit}
\end{eqnarray}
where $p_0$ is a floating normalization parameter, $p_1$ is a fitting parameter related 
to the radius by $R = \sqrt{-6 p_1}$, parameters for higher order terms ($p_{i}$ with $i > 1$) 
are free fitting parameters, and $N$ is defined by the user.

\paragraph{Multi-parameter rational-function of $Q^2$}
The fitter of the multi-parameter rational-function of $Q^2$ is expressed as
\begin{eqnarray}
f_{\textrm{rational}}(Q^2) &=& p_0 G_E(Q^2) = p_0 \frac{1 + \sum\limits_{i=1}^{N} p^{(a)}_{i} Q^{2i}}{1 + 
\sum\limits_{j=1}^{M} p^{(b)}_{j} Q^{2j}} , \label{poly-ratio-Q2-fit}
\end{eqnarray}
where $p_0$ is a floating normalization parameter, $p^{(a)}_{i}$ and $p^{(b)}_{j}$ are free fitting parameters, 
and radius can be found as $R = \sqrt{6(p^{(b)}_1-p^{(a)}_1)}$. The orders $N$ and $M$ are defined by the user.

\paragraph{CF expansion}
The CF expansion fitter is expressed as \cite{Sick_2003gm}
\begin{eqnarray}
f_{\textrm{CF}}(Q^2) &=& p_0 G_E(Q^2) = p_0 \frac{1}{1+\frac{p_1 Q^2}{1+\frac{p_2 Q^2}{1+ \cdots}}} , \label{cf-fit}
\end{eqnarray}
where $p_0$ is a floating normalization parameter, $p_i$ ($i>0$) are free fitting parameters, and $R = \sqrt{6 p_1}$. 
The user can define the maximum $i$ of the expansion.

\paragraph{Multi-parameter polynomial-expansion of $z$}
The $z$-transformation is expressed as \cite{Lee_2015jqa}
\begin{eqnarray}
z &=&  \frac{\sqrt{T_c + Q^2} - \sqrt{T_c - T_0}}{\sqrt{T_c + Q^2} + \sqrt{T_c - T_0}}, \label{z-trans}
\end{eqnarray}
where $T_c = 4 m_{\pi}^2$, $m_{\pi}$ is set to be 140 MeV (close to the $\pi^0$ mass as in Ref. \cite{Lee_2015jqa}), 
and $T_0$ is a free parameter representing the point mapping onto $z=0$ ($T_0$ is set to $0$ in this study).
With the new variable $z$, $G_E$ can be parameterized as
\begin{eqnarray}
f_{\textrm{polyz}}(Q^2) &=& p_0 G_E(Q^2) = p_0 \left(1 + \sum\limits_{i=1}^N p_{i} z^{i} \right) , \label{poly-z-expan-fit}
\end{eqnarray}
where $p_0$ is a floating normalization parameter, $p_1$ is a fitting parameter related to the radius 
by $R = \sqrt{-3 p_1/ 2 T_c}$, $p_{i}$ are free fitting parameters, and $N$ is defined by the user.

%
%
\section{\label{tests}
Tests over full range of the PRad kinematics}

\begin{figure*}[htb]
\includegraphics[width=0.9\textwidth]{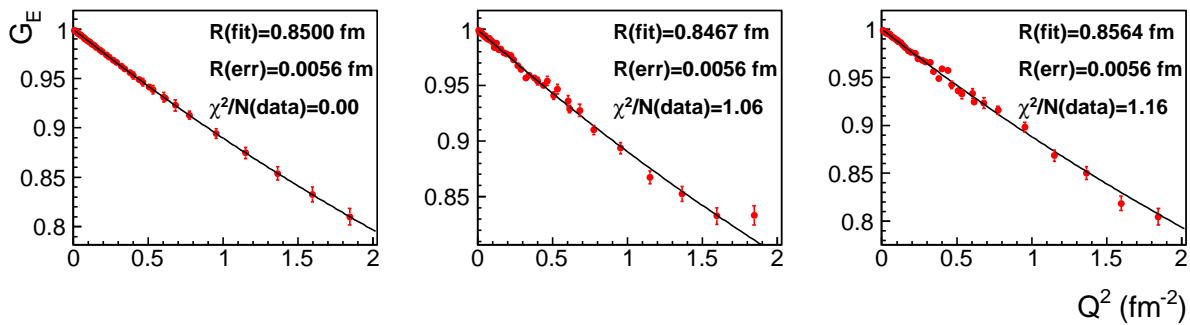}
\caption{\label{dipole_3_fits}(color online).  Left panel shows pseudo-data generated with dipole functional 
form without fluctuations and the middle and right panel shows results with fluctuations [$R(\textrm{input}) = 0.85$ fm].
The fitting result [$R(\textrm{fit})$], fitting 
uncertainty [$R(\textrm{err})$] and $\chi^2$ per data point [$\chi^2/N{\textrm{(data)}}$] 
are presented in each panel.}
\end{figure*}

We tested functions over the expected $Q^2$ range for the PRad experiment of
$3\times 10^{-4} <  Q^2 < 0.072$ GeV$^{2}$, using bin-by-bin
random uncertainties from 0.02$\%$ to 1.1$\%$.  The
exact values used can be found online and are denoted as bin set one~\cite{Bin_sets}.
As an example, Fig.~\ref{dipole_3_fits} shows pseudo-data generated 
with the dipole generator [$R(\textrm{input})=0.85$ fm] in the PRad binning fit using
the dipole fitter. 
In the first panel, no fluctuation is added to the central values of $G_E$ while the
other panels show two of the many possible outcomes of adding random fluctuations to the pseudo-data with the same input parameters.
As one might expect, when there is no fluctuation, the fit curve goes through all the pseudo-data points 
perfectly and the input $R$ value is obtained.  However, when there are fluctuations, the results of the 
fit can differ from the input.

In order to determine the distribution of possible outcomes, one needs to generate many sets of pseudo-data
and perform fits for each set.  This is done using the following procedure:

\paragraph{\label{step_g}Generation}
First, one $G_E$ model is used to generate pseudo-data (using the generator), at 
the bin centers of $Q^2$ that the user inputs into the program. 

\paragraph{\label{step_fa}Fluctuation-adding}
Next, bin-by-bin and overall fluctuations are added to the $G_E$ vs. $Q^2$ tables 
in a random manner (using the fluctuation-adder), to mimic the real data. 
The bin-by-bin uncertainties are taken from the bin-set file, and an overall scaling 
uncertainty of $5\%$ (far larger than expected in the PRad result) is added in 
the tests to show that this method works even if there is such a big scaling uncertainty.

\paragraph{\label{step_f}Fitting}
Finally, the $G_E$ vs. $Q^2$ tables are fit with a number of functional forms 
(using the fitting library) to extract $R$ from the pseudo-data with fluctuations.

The steps of generation, fluctuation-adding and fitting are repeated 150,000 
times for each combination of generator and fit function. 
The 150,000 fitting results of $R(\textrm{fit})$ for each combination comprise a distribution with a central 
value $R(\textrm{mean})$ and a root-mean-square (RMS) width.
As the fitting uncertainty of $R$, determined by Minuit (for each of the 150,000 fits) 
is very close to the RMS width of the $R(\textrm{fit})$ distribution, we will use 
the RMS values to represent the one-$\sigma$ fitting-uncertainty.

\subsection{\label{fit_strong_model}Fits with simple-function models}

\begin{figure*}[htb]
\includegraphics[width=0.9\textwidth]{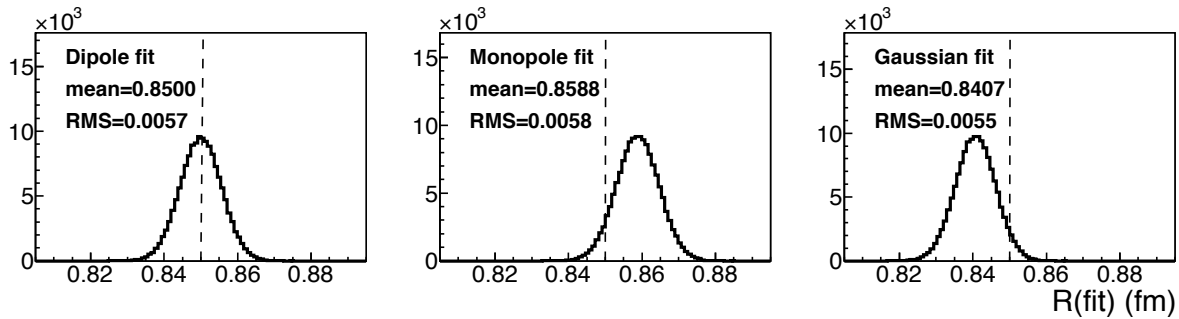}
\caption{\label{dipole_gen_d_m_p_fits}  Dipole, monopole and Gaussian fits of 
pseudo-data tables generated with the dipole functional form and added fluctuations.
The dashed line indicates the value of the input radius of 0.85~fm.}
\end{figure*}

Fig.~\ref{dipole_gen_d_m_p_fits} shows the $R(\textrm{fit})$ distributions of the dipole, monopole and Gaussian 
fits when the dipole generator is used [$R(\textrm{input}) = 0.85$ fm].
It is observed that when the dipole fitter is used, $R(\textrm{mean}) \approx R(\textrm{input})$, 
but when the monopole or Gaussian fitter is used, $R(\textrm{mean})$ significantly deviates from $R(\textrm{input})$.

\begin{figure*}[htb]
\includegraphics[width=0.8\textwidth]{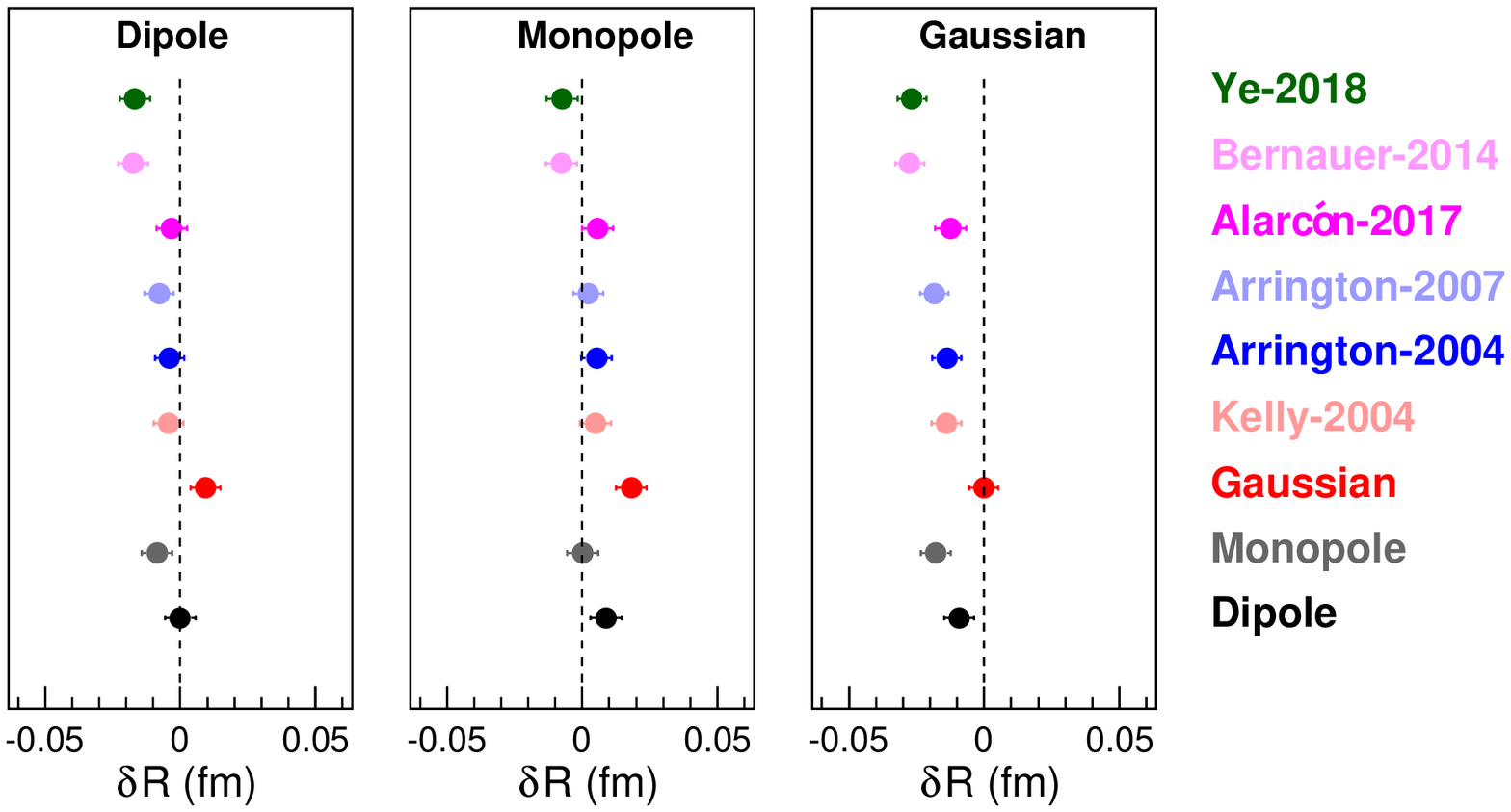}
\caption{\label{dR_dmg_3_plots_row}(color online). Dipole, monopole and Gaussian 
fits of pseudo-data tables generated with nine models.}
\end{figure*}

Fig.~\ref{dR_dmg_3_plots_row} summarizes the fitting results using the dipole, monopole and Gaussian fitter, 
respectively, when nine generators covering nine of the $G_E$ models describe in section~\ref{generator}.
It is clear that the simplest functional forms are 
not able to provide a robust extraction of $R$ over the full kinematic range of PRad bins, 
since for various input $G_E$ models, the fitting uncertainty ($\sigma$) is smaller than the 
size of the bias [$\delta R = R(\textrm{mean}) - R(\textrm{input})$].   We note that in this
type of statistical analysis, bias is simply the mean offset from the input value and is
not meant as a pejorative term.   In fact, it is the trade-off between bias and variance that
is at the heart of machine learning algorithms~\cite{Hastie:2009}.

\subsection{\label{fit_poly_Q2}Fits with polynomial expansions of $Q^2$}
\begin{figure*}[htb]
\includegraphics[width=\textwidth]{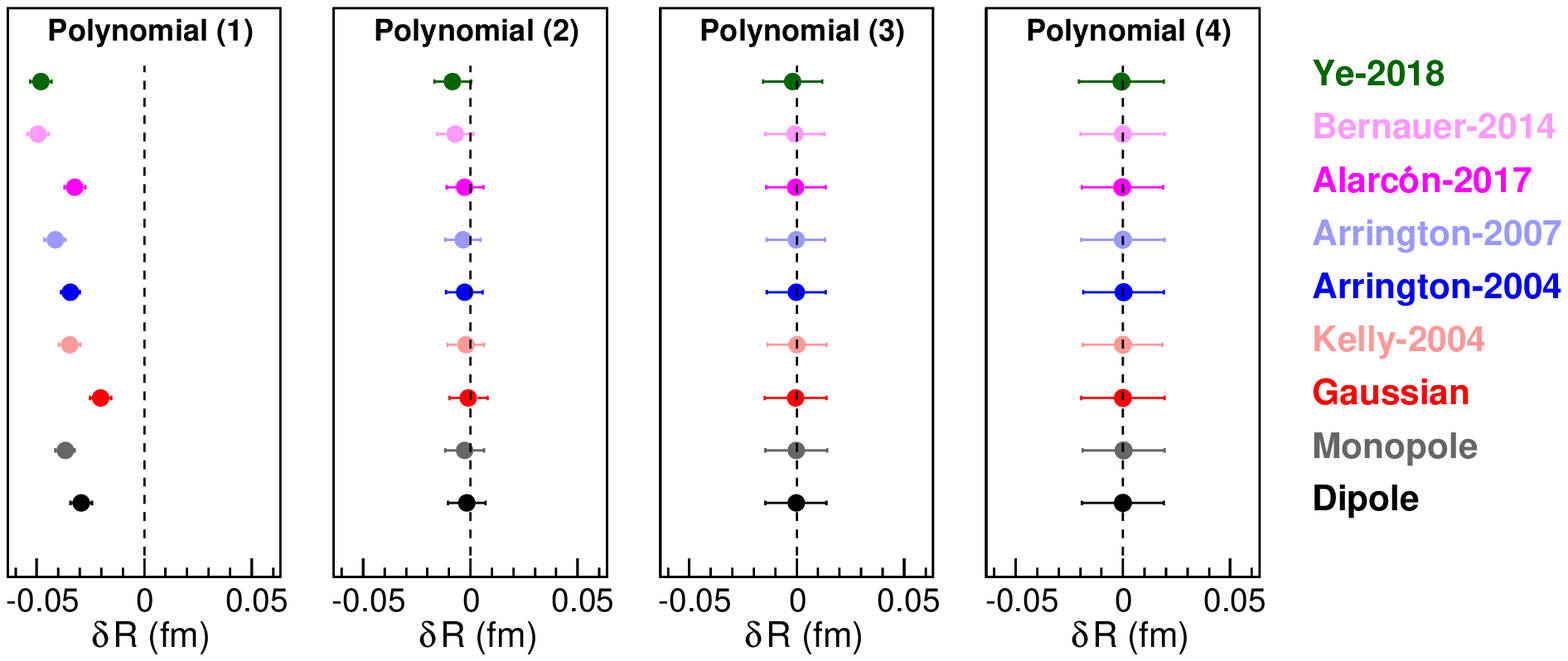}
\caption{\label{dR_polyQ_4_plots_row}(color online).  Polynomial-expansion 
fits of pseudo-data tables generated with nine models.}
\end{figure*}

Polynomial expansions have been widely used to fit $G_E$ vs. $Q^2$ data,
though concerns have been raised about extrapolating with
polynomial functions~\cite{Griffioen_2015hta, Higinbotham_2015rja, Horbatsch_2015qda, Sick_2017aor, Kraus_2014qua}.
Fig.~\ref{dR_polyQ_4_plots_row} summarizes the fitting results using the polynomial-expansion 
fitter with $N=1$, $2$, $3$ and $4$, using again the nine generators covering various types of $G_E$ models.

For the full range of the PRad data, the first-order polynomial fitter is clearly not robust
as $|\delta R| > \sigma$ for all the input $G_E$ models.
The second-order polynomial is marginally robust, since $|\delta R| \approx \sigma$ 
is found for models Bernauer-2014 and Ye-2018, while $|\delta R| < \sigma$ is found 
for the others.  The third and fourth order polynomial fits seem to be robust with
$|\delta R| < \sigma$ for all the input $G_E$ models, but with a 
significantly larger $\sigma$.

We observe that when the order of expansion is too low ($N=1$), $R(\textrm{mean})$ 
is systematically and significantly smaller than $R(\textrm{input})$ for all the generators 
used in the tests.  The polynomial (1) fit shows high bias and a low variance.
When higher orders are included ($N=2$, $3$ and $4$), $R(\textrm{mean})$ gets 
closer to $R(\textrm{input})$, regardless of the type of generator.
At the same time, 
as the number of parameters increases the fitting uncertainties increases, showing low bias with
high variance.
The optimal choice of $N$ depends on the $Q^2$ range, the distance between bin centers and
the uncertainty level in the data table.
This clearly illustrates the trade-off between bias and variance and the need to balanced
them when fitting.  Some efforts have been taken to build algorithms that automatically 
and systematically determine the 
proper order $N$ when fitting certain data~\cite{Higinbotham_2015rja,Hayward:2018qij}.

\subsection{\label{fit_poly_rato_Q2}Fits with rational functions of $Q^2$}
\begin{figure*}[htb]
\includegraphics[width=\textwidth]{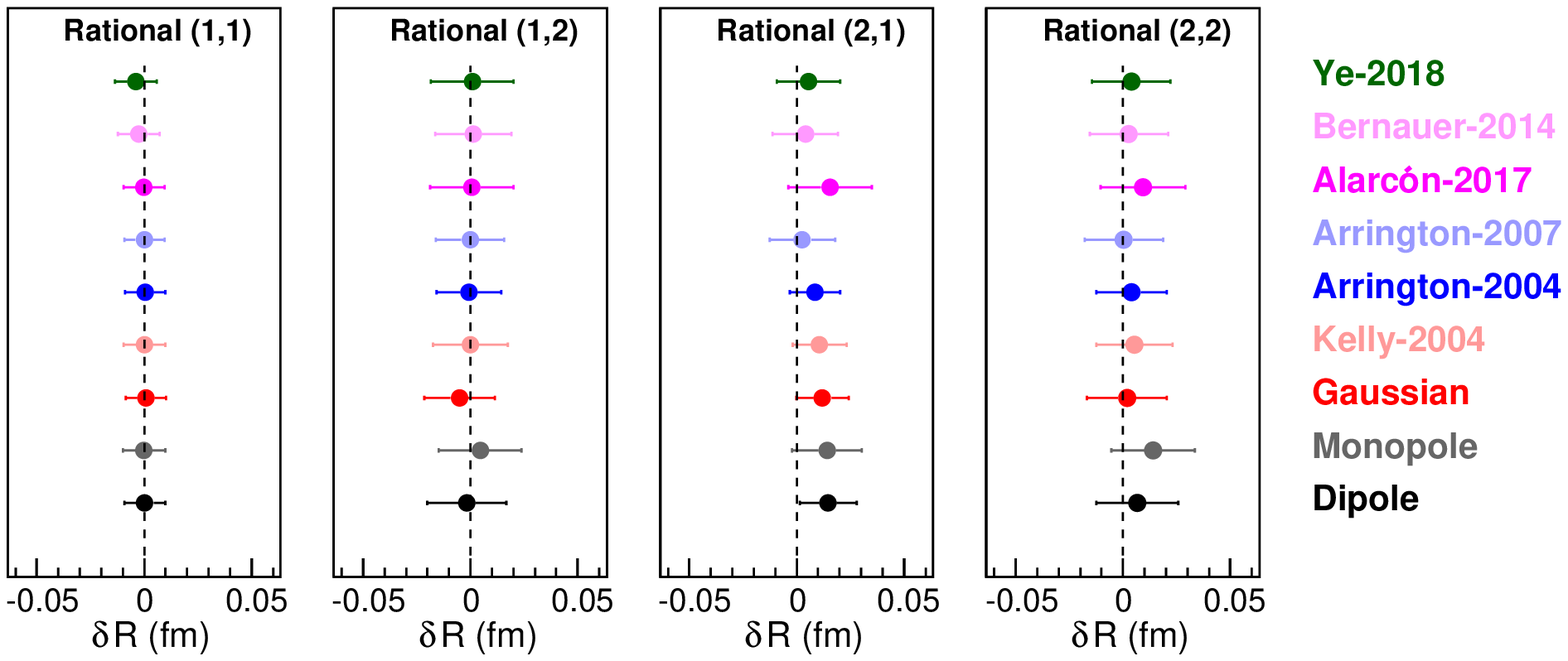}
\caption{\label{dR_rational_4_plots_row}(color online).  Rational function fits of 
pseudo-data tables generated with nine models.}
\end{figure*}

Rational functions are also widely used to fit 
$G_E$ vs. $Q^2$ data, such as in Refs.~\cite{Kelly_2004hm,Arrington_2003qk,Venkat_2010by,Puckett:2017flj}.
Fig.~\ref{dR_rational_4_plots_row}  summarizes the fitting results using the rational-function 
fitter with $(N, M)=$ ($1$, $1$), ($1$, $2$), ($2$, $1$) and ($2$, $2$), using the same
nine generators. 
In these tests, the rational-function fitter ($N,M)$=$(1,1)$ extracts $R$ robustly ($\delta R < 0.42\sigma$) 
regardless of the model parameterization in the generator.
It also has the lowest fitting uncertainty among these four rational-function parameterizations.
The higher order
rational-function fitters, are also robust ($|\delta R| < \sigma$ for all input $G_E$ models) but
a have significantly larger fitting uncertainties.

\subsection{\label{fit_CF_Q2}Fits with Continued Fractions}
Using Continued Fraction, CF, expansions to fit $G_E$ vs. $Q^2$ data was proposed and 
applied to the world data by Sick in 2003~\cite{Sick_2003gm}.
This work also included tests and discussions regarding fitting pseudo and real data with CF expansions.

\begin{figure*}[htb]
\includegraphics[width=\textwidth]{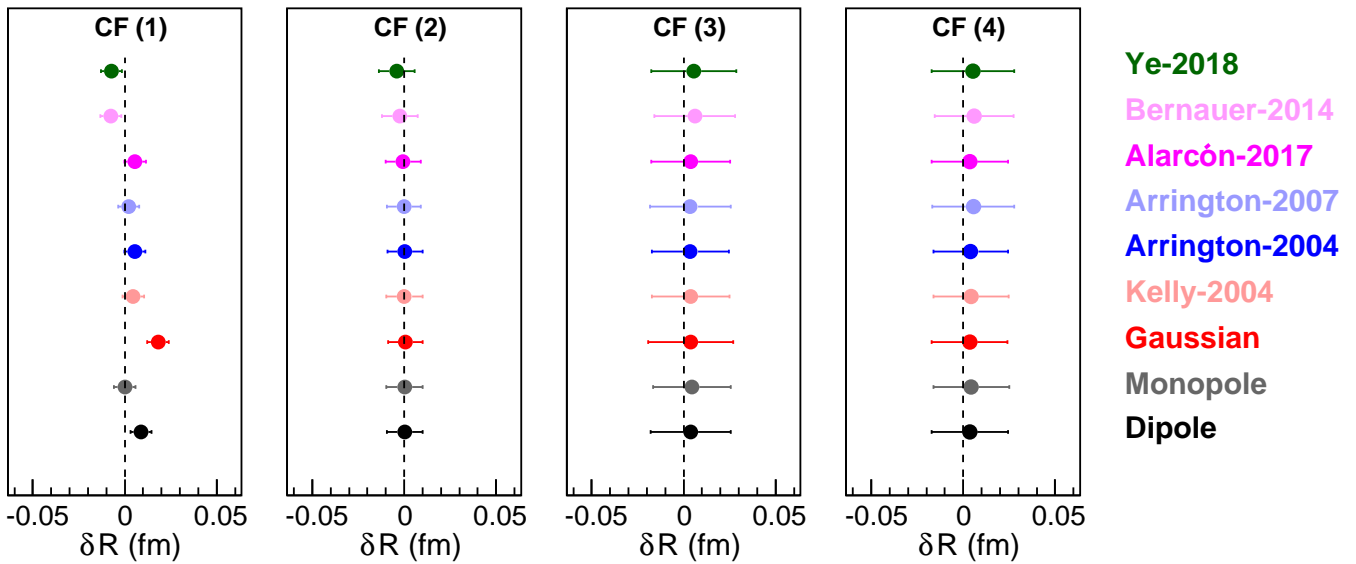}
\caption{\label{dR_cf_4_plots_row}(color online). CF fits of pseudo-data tables generated with nine $G_E$ models.}
\end{figure*}

Fig.~\ref{dR_cf_4_plots_row} summarizes results using the CF fit at order 1, 2, 3 and 4, 
with the same nine generating models.
In these tests (using PRad binning), the second order CF is robust: $|\delta R| < \sigma$, 
regardless of the parameterizations in the generator, and the fitting uncertainties are small.
Higher-order CF fitters, while robust ($|\delta R| < \sigma$ for all input $G_E$ models), 
have significantly larger fitting uncertainties.

\subsection{\label{fit_poly_z}Fits with polynomial expansions of $z$}
Using polynomial expansion of $z$ instead of $Q^2$ is another option to extract $R$.  Here Eq.~(\ref{z-trans})
is used to transform $Q^2$ to $z$.

\begin{figure*}[htb]
\includegraphics[width=\textwidth]{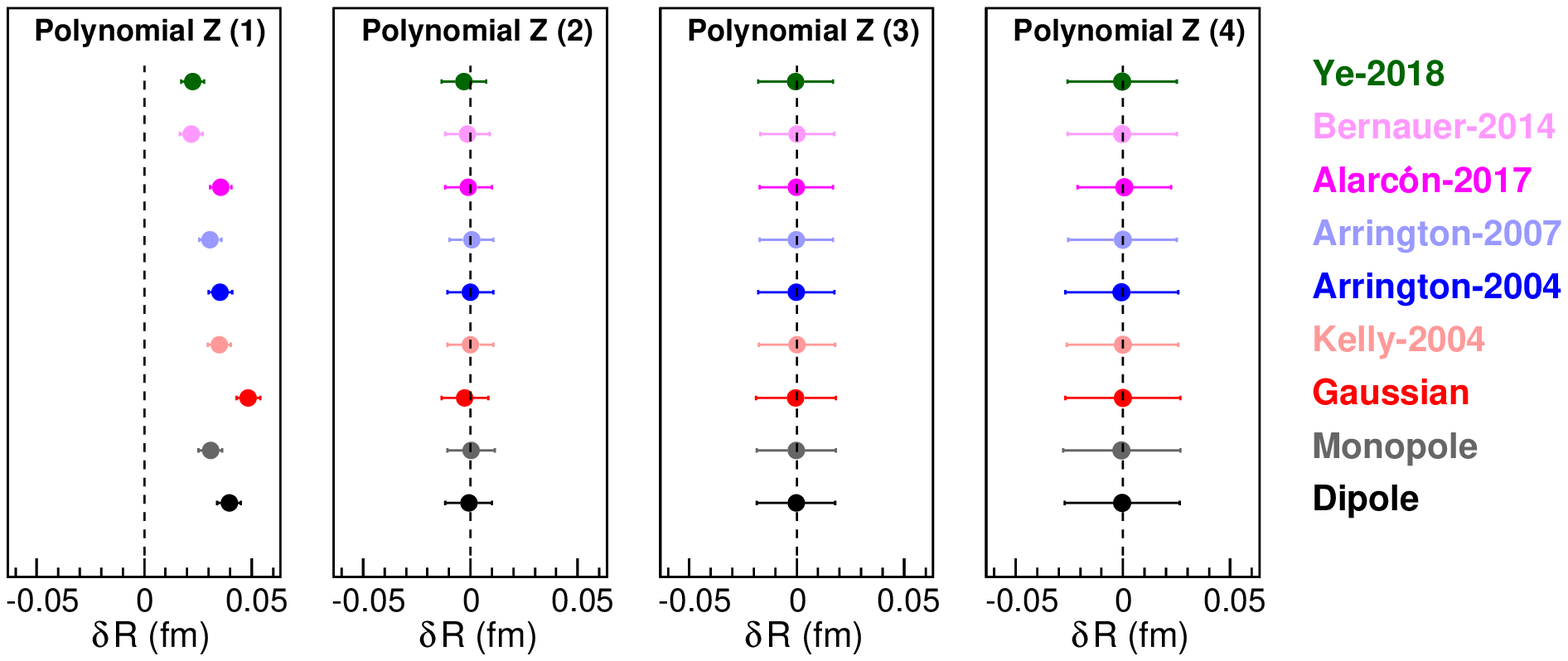}
\caption{\label{dR_polyZ_4_plots_row}(color online). Polynomial-expansions-of-$z$ fits of 
pseudo-data tables generated with nine models.}
\end{figure*}

Fig.~\ref{dR_polyZ_4_plots_row} summarizes the fitting results using polynomial expansions of $z$ 
with $N=1$, 2, 3 and 4, using the nine generator functions.
When $N=1$, $R(\textrm{mean})$ is systematically and significantly larger 
than $R(\textrm{input})$ for all the generators used in the tests, opposite to the systematically smaller $R(\textrm{mean})$
for the polynomial (1) fits in $Q^2$.
Again, as higher-order terms are included in the polynomial expansion of $z$, 
the bias is reduced though sigma increases.
The polynomial expansion of $z$ with $N=2$ is clearly the best as it is robust and
has a small fitting uncertainty.

%
%
\section{\label{rb_in_diff_bin_cond}
Tests over low $Q^2$ subsets of the PRad kinematics}

In this section, we consider extracting $R$ using 
only low $Q^2$ subsets of the PRad kinematics.   As the amount of
data in the extremely low $Q^2$ ranges are quite limited, it is 
easy to overfit the data and cause huge variances, so only the fit
functions that give reasonable results are shown.

\begin{figure*}[htb]
\includegraphics[width=\textwidth]{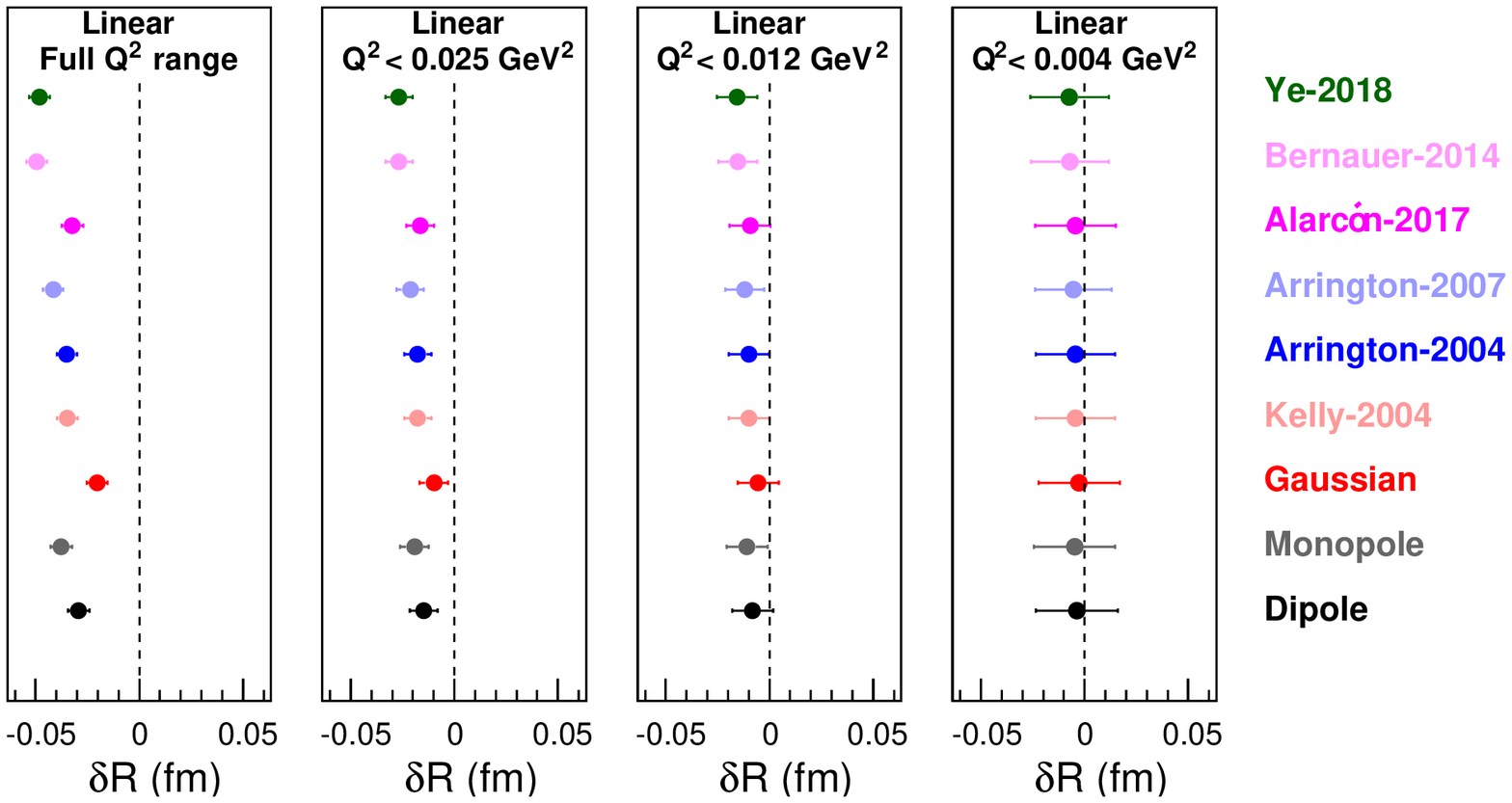}
\caption{\label{dR_lowQ2_series1}(color online).   While low $Q^2$ linear fits have a large bias when used full range;
as the range in $Q^2$ is decrease, the bias decreases while the sigma
increases.}
\end{figure*}

\begin{figure*}[htb]
\includegraphics[width=\textwidth]{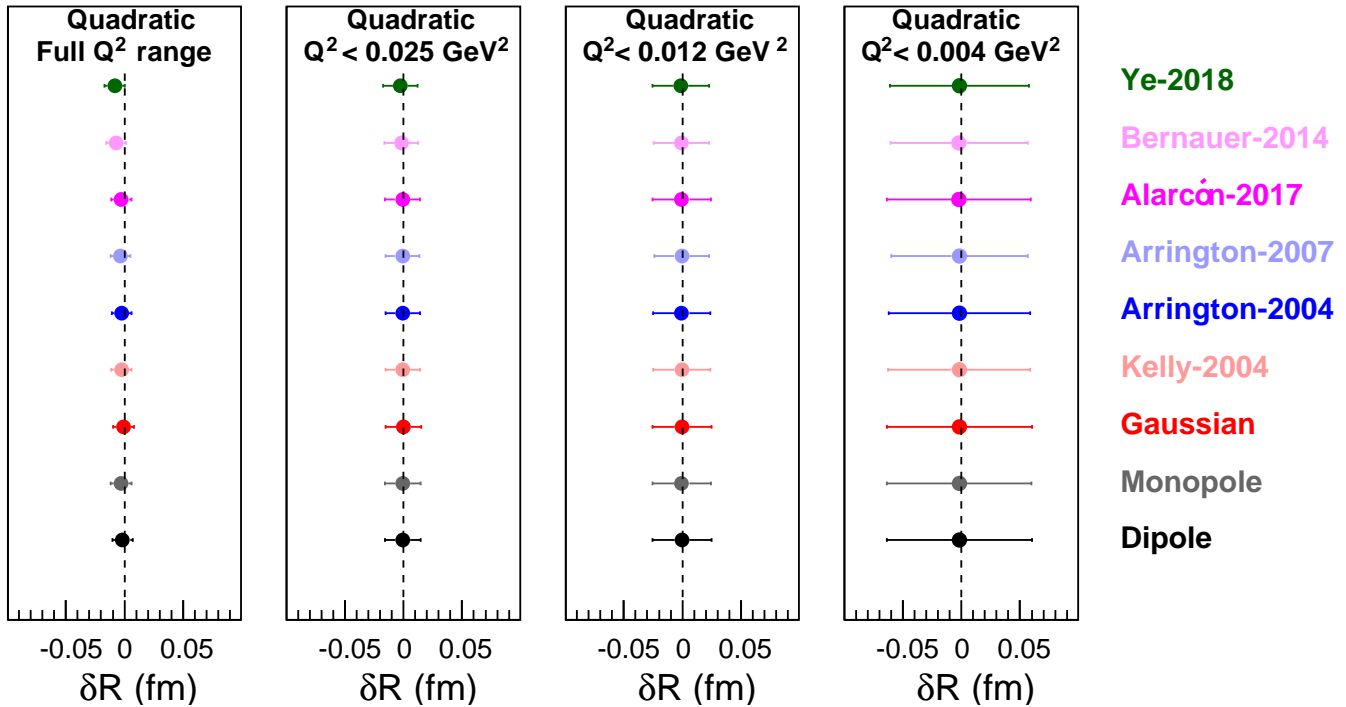}
\caption{\label{dR_lowQ4_series1}(color online).   A series of quadratic fits 
over low $Q^2$ ranges showing that for the smallest ranges the variance gets
huge and the function overfits the data.}
\end{figure*}

\begin{figure*}[htb]
\includegraphics[width=\textwidth]{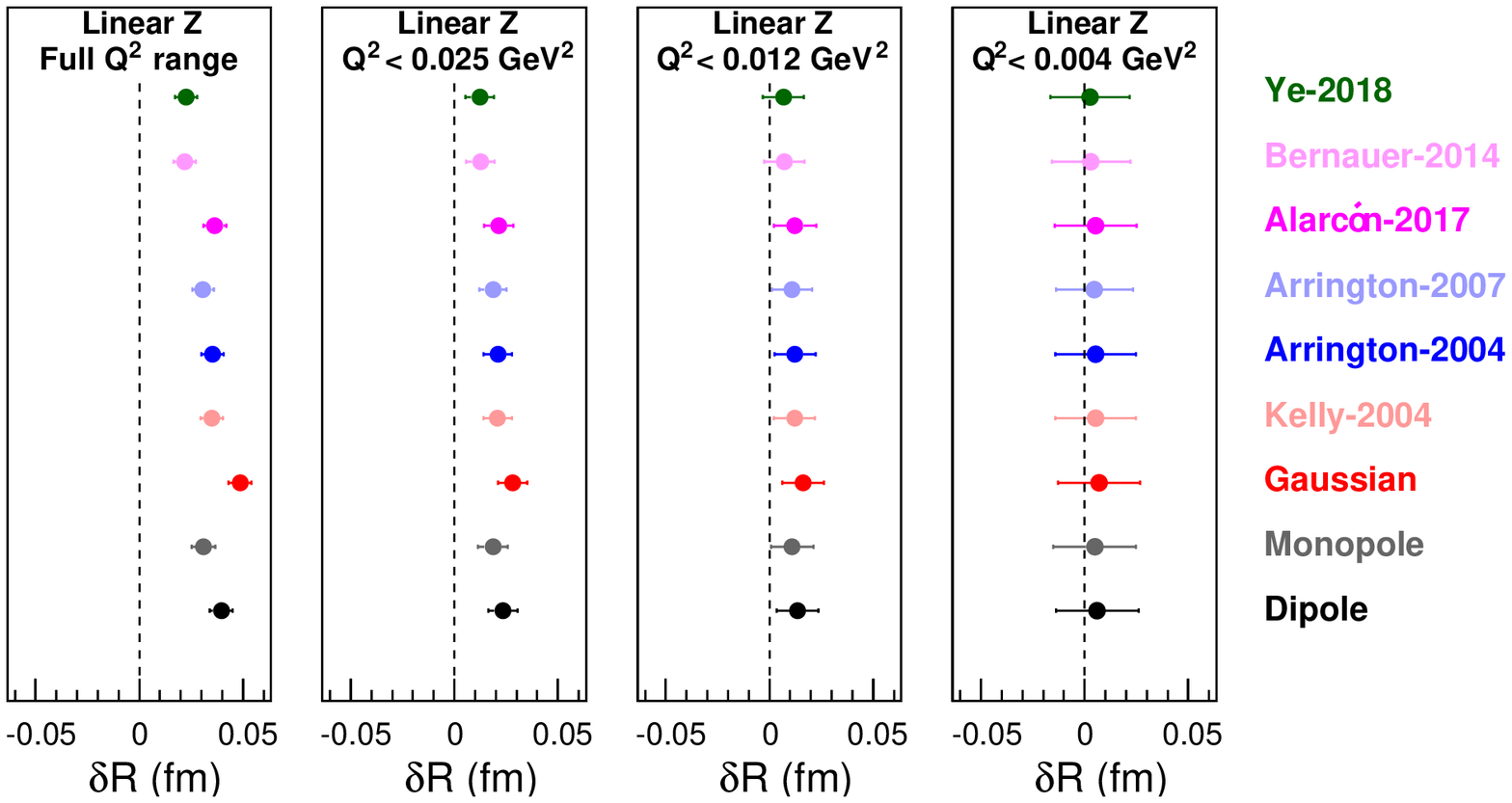}
\caption{\label{dR_lowz1_series1}(color online).  A series of linear fits
in z; again showing a trade-off between bias and variance as the upper limit in
$Q^2$ is decreased.}
\end{figure*}

Fig.~\ref{dR_lowQ2_series1}, we show linear fits, first
order polynomial, for various range of $Q^2$.    For the full $Q^2$ range,
the linear fit produces a large bias and has a poor
residual, while as the range of the data is reduced,
the linear fit becomes robust.
In Fig.~\ref{dR_lowQ4_series1}, we show
that though the quadratic fits works reasonably well over the full
range of the expect data, as the range is restricted the quadratic
function quickly starts to overfit the 
data and has a much larger variance than the linear fits.

Repeating the linear fits in z instead of $Q^2$, again has the interesting
effect that the bias has the opposite sign, as shown in Fig.~\ref{dR_lowz1_series1}.
Nevertheless, as the $Q^2$ range is decreased, the bias is reduced
and the linear fits in both $G_E$ vs. $Q^2$ and $G_E$ vs. z agree.
This emphasizes that merely
transforming to z does not eliminate the problem
of selecting the appropriate functional form to fit the data.
Though the results do show that for very low $Q^2$, linear fits 
in z and $Q^2$ should agree.

%
%
\section{\label{discussion}Discussion}

Choosing the appropriate fit function to extract $R$ for a given set of data
depends on both the $Q^2$ ranges and bin-by-bin uncertainties, thus, the 
the choice of appropriate fitting function(s) needs to be determined on a 
case-by-case basis.  In fact, Figs.~\ref{dR_lowQ2_series1} and \ref{dR_lowQ4_series1} show
that it is possible that simpler 
fitters are robust and have smaller $\sigma$ then more complex functions
when focusing low-$Q^2$ subsets of data.   It is therefore imperative to define the 
criteria for selecting functions for extracting $R$.

A standard way of quantifying goodness of fit for this type of study where the true
values are known is to consider both the bias and variance~\cite{Hastie:2009} using Root
Mean Square Error, RMSE, where
\begin{eqnarray}
\textrm{RMSE} &=& \sqrt{\textrm{bias}^2 + \sigma^2} . \label{mse-eq}
\end{eqnarray}
In this study, $\delta R$ is the bias and $\sigma$ is represented by the RMS value
of the fitting results.  

\begin{figure*}[htb]
\includegraphics[width=\textwidth]{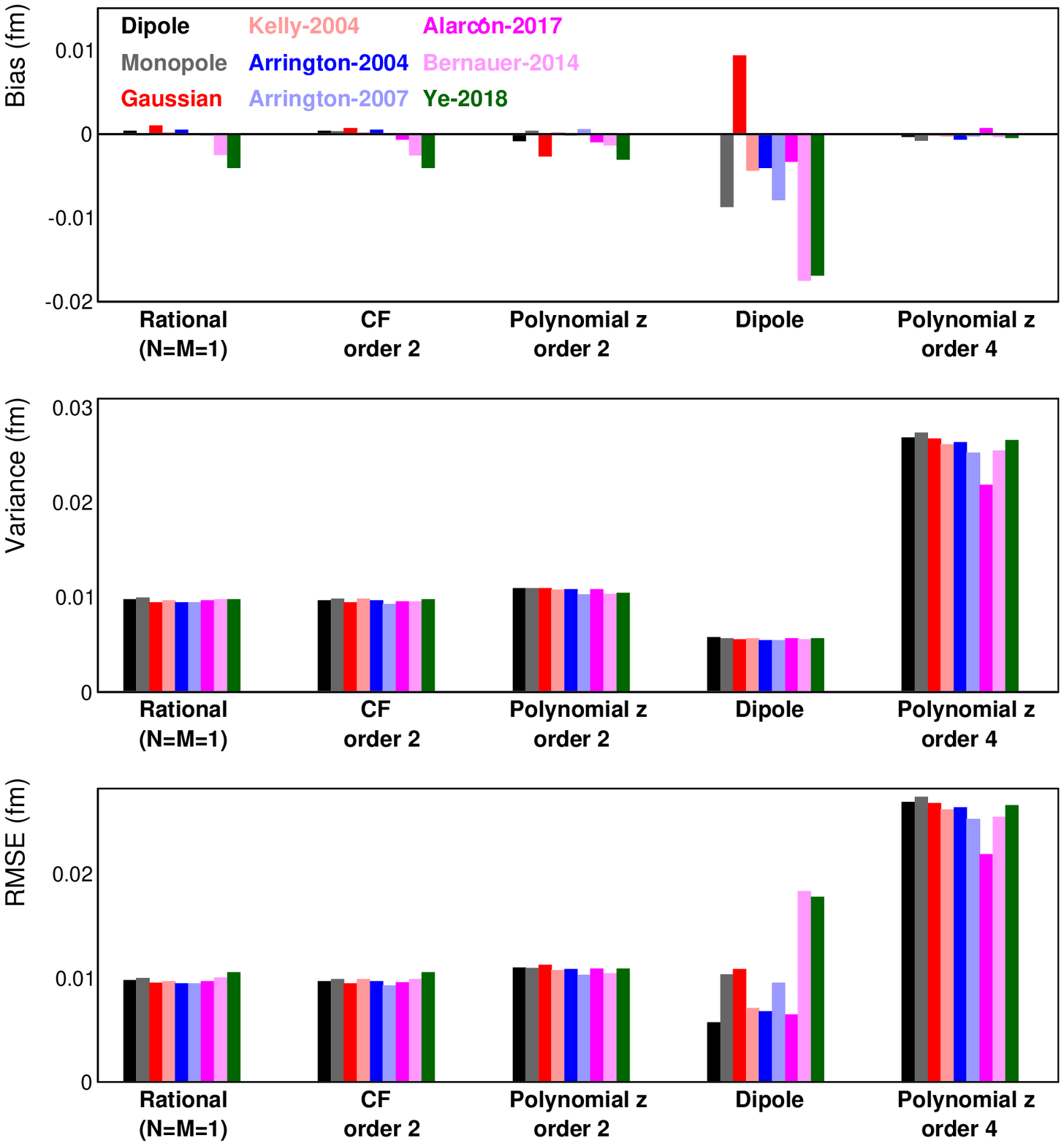}
\caption{\label{bias_sigma_mse}(color online). Bias ($\delta R$), variance ($\sigma$) and RMSE 
of the rational ($N=1, M=1$), 
CF (second order), polynomial expansion of $z$ ($N=2$), dipole and polynomial expansion of $z$ ($N=4$) 
fitters. The last two fitters represent typical cases of under-fit (large bias and small variance) 
and over-fit (small bias and large variance), respectively.  The bias, $\sigma$ and RMSE 
values of nine $G_E$ generators with the fitters presented by the nine colored columns 
in the corresponding to dipole, 
monopole, Gaussian, Kelly-2004, Arrington-2004, Arrington-2007, Alarc{\'o}n-2017, Bernauer-2014 and Ye-2018 
respectively.}
\end{figure*}

Fig.~\ref{bias_sigma_mse} summarizes the bias, $\sigma$ and RMSE values for
the three good fitting functions, one of the large-bias fitters (dipole) 
and one of the large-variance fitters [polynomial expansion of $z$ ($N=4$)]
for the full range of the expected PRad data.
The RMSE values of the three good fitters are similar for all generating functions.  
The RMSE values of the large-bias fitter, though 
smaller than those of the three good fitters on average, have large variations when different generators 
are used, which indicates that the fitter is not robust.
The RMSE values of the large-variance fitter are significantly larger than those of the good fitters, 
which indicates that too many parameters were used.

\begin{figure}[htb]
\includegraphics[width=\columnwidth]{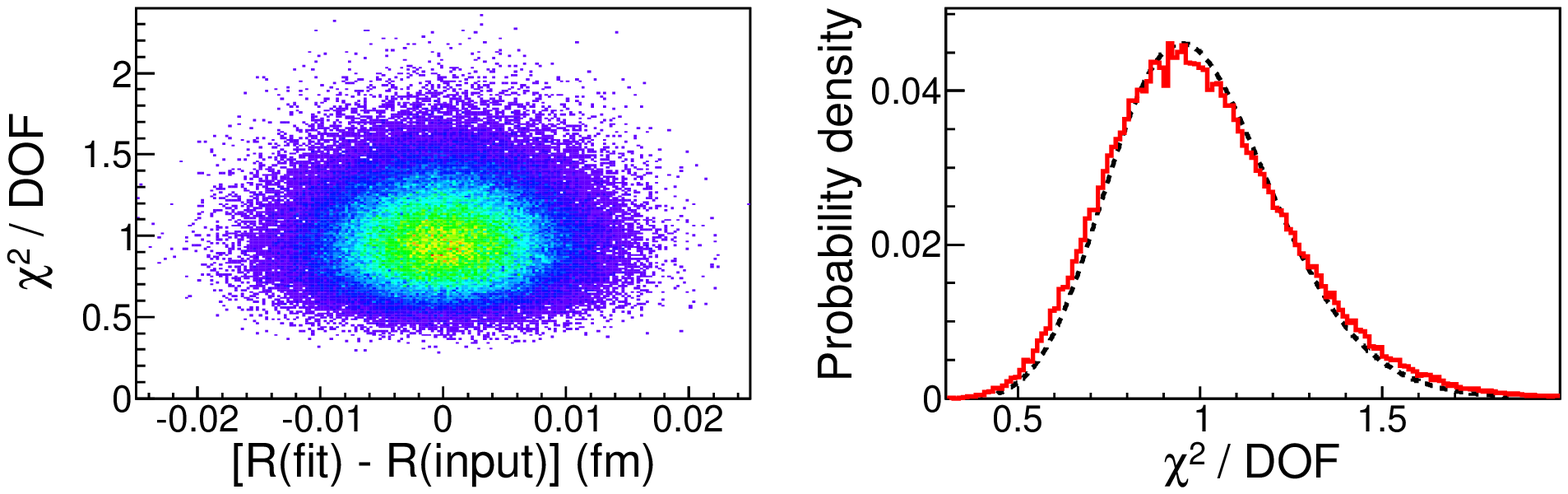}
\caption{\label{dipole_gen_chi2_vs_dR}(color online).  The left panel presents the correlation 
between $\chi^2/ \textrm{DOF}$ and [$R(\textrm{fit}) - R(\textrm{input})$], when both the 
generator and the fitter use the dipole functional form. The black dashed curve in the right 
panel presents the ideal probability density function of $\chi^2 / \textrm{DOF}$ distribution, 
and the red curve is from the numerical tests.}
\end{figure}

The $G_E$ values in real data inevitably have some fluctuations around the true central value
due to statistical and systematic uncertainties.
To test if these fluctuations have been correctly accounted for in the tests herein, we check
the distribution of our results against an ideal probability density function.
The left panel of Fig.~\ref{dipole_gen_chi2_vs_dR} 
shows the correlation between the $\chi^2$ per degree of freedom (DOF) and [$R(\textrm{fit}) - R(\textrm{input})$], 
where $\textrm{DOF}= N(\textrm{data}) -2$, and $N(\textrm{data})$ is the number of data points in the $G_E$ vs. $Q^2$ table. 
The black curve in the right panel of Fig.~\ref{dipole_gen_chi2_vs_dR} is the ideal probability density function 
of $\chi^2 / \textrm{DOF}$ distribution, and the red curve is from the numerical tests. The good agreement
between these two curves indicates that the tests work as expected \cite{Sirca2012, Sirca2016}.
In this figure, both the generator and the fitter use the dipole functional form, though similar results
were achieved for all functional forms.

The value of R extracted from a known generator can vary due to fluctuations even if the
$\chi^2 / \textrm{DOF}$ is reasonable.
Additionally, a good $\chi^2$ value is not sufficient to determine if the corresponding 
fit can extrapolate the radius properly.  From a purely mathematical point of view, this can be understood as the 
difference between a good interpolating function, valid over the range of the
data, and a functional form that can be used to extrapolate beyond the range of the data.

For, with real data, with one only a single data set and an unknown functional form, it is not possible to 
know exactly how much the fluctuations  affect the $R$ extraction unless $R$ is already known
precisely.  On the other hand, one can make use of the statistical bootstrap which uses sampling
with replacement to produce multiple data tables from a single set of data~\cite{Efron:1979,Efron:1987}.   
While the bootstrap won't affect the mean, it allows determination of the uncertainty distribution 
from the data itself.

Theoretical models, such as Alarc{\'o}n-2017, can also be 
used to help with fitting experimental data and extracting $R$.  For example, one can use theory
to constrain high order moments and achieve a smaller fit uncertainties~\cite{Horbatsch:2016ilr}, though 
these approaches inevitably introduce theory dependence to the $R$ extraction.
Theory dependence has been avoided in this study and we have demonstrated that 
the robust fitting functions are able to extract $R$ by
using relatively simple functions and
treating the higher order moments as nuisance parameters.
Of course, a pure mathematical extraction, such as
demonstrated in this paper, and a valid 
nuclear theory extraction should give the same radius within errors. 

%
%
\section{Summary}
We have created an expandable framework to search for functional forms
that can reliably extract the proton radius using
pseudo-data generated from a wide variety of $G_E$ models.
As a pertinent example, we have applied this framework to the
expected range and uncertainty of the upcoming PRad data.

We find that for the full range of the data, the $(N=M) = (1,1)$ rational function, the two parameter 
continued fraction, and the second order polynomial expansion in $z$ can all robustly extract 
the correct radius with small $\sigma$ regardless of the input pseudo-data generating function.
By restricting data to the lowest $Q^2$ ranges, it is also possible to extract the radius
using a linear function though this yields a larger uncertainty
than when using the full range.
We also note that functions with a good $\chi^2$ do not necessarily
extrapolate well beyond data; thus $\chi^2$ alone cannot be used to determine which functions 
can robustly extract the proton radius.

\begin{acknowledgments}
This work is supported by the U.S. Department of Energy, Office of Science, Office of Nuclear Physics 
under contract DE-AC05-060R23177 and supported in part by the U.S. Department of Energy under 
Contract No. DE-FG02-03ER41231, Thomas Jefferson National 
Accelerator Facility and Duke University.  We acknowledge the helpful discussions with 
Jose Alarc{\'o}n, John Arrington, Carl Carlson, Keith Griffioen, Ingo Sick, Simon {\v{S}}irca, 
Christian Weiss and Zhihong Ye.
We also acknowledge the support and encouragement from Robert D. McKeown.
\end{acknowledgments}

\bibliography{reference}

\end{document}